\documentclass[%
aip,
rsi,%
amsmath,amssymb,
reprint,%
]{revtex4-1}

\usepackage{graphicx}% Include figure files
\usepackage{dcolumn}% Align table columns on decimal point
\usepackage{bm}% bold math
\begin{document}
	
	\title{Computational metrics and parameters of an injection-locked large area semiconductor laser for neural network computing}
	
	\author{Anas Skalli}
	\email{anas.skalli@femto-st.fr}
	\affiliation{FEMTO-ST Institute/Optics Department, CNRS \& University Bourgogne Franche-Comt\'e, \\15B avenue des Montboucons,
		Besan\c con Cedex, 25030, France.}%Lines break automatically or can be forced with \\

	\author{Xavier Porte}%
	\affiliation{FEMTO-ST Institute/Optics Department, CNRS \& University Bourgogne Franche-Comt\'e, \\15B avenue des Montboucons,
	Besan\c con Cedex, 25030, France.}%Lines break automatically or can be forced with \\

	\author{Nasibeh Haghighi}%
	\affiliation{echnical University Berlin/Institute of Solid-State Physics and the Center of Nanophotonics, Hardenbergstra{\ss}e 36, D-10632 Berlin, Germany.}%Lines break automatically or can be forced with \\

	\author{Stephan Reitzenstein}%
	\affiliation{echnical University Berlin/Institute of Solid-State Physics and the Center of Nanophotonics, Hardenbergstra{\ss}e 36, D-10632 Berlin, Germany.}%Lines break automatically or can be forced with \\

	\author{James A. Lott}%
	\affiliation{echnical University Berlin/Institute of Solid-State Physics and the Center of Nanophotonics, Hardenbergstra{\ss}e 36, D-10632 Berlin, Germany.}%Lines break automatically or can be forced with \\

	\author{D. Brunner}
	\affiliation{FEMTO-ST Institute/Optics Department, CNRS \& University Bourgogne Franche-Comt\'e, \\15B avenue des Montboucons,
		Besan\c con Cedex, 25030, France%\\This line break forced% with \\
	}%

	\date{\today}% It is always \today, today,
	
	\begin{abstract}
		
Artificial neural networks have become a staple computing technique in many fields. Yet, they present fundamental differences with classical computing hardware in the way they process information. Photonic implementations of neural network architectures potentially offer fundamental advantages over their electronic counterparts in terms of speed, processing parallelism, scalability and energy efficiency. Scalable and high performance photonic neural networks (PNNs) have been demonstrated, yet they remain scarce. In this work, we study the performance of such a scalable, fully parallel and autonomous PNN based on a large area vertical-cavity surface-emitting laser (LA-VCSEL). We show how  the performance varies with different physical parameters, namely, injection wavelength, injection power, and bias current. Furthermore, we link these physical parameters to the general computational measures of consistency and dimensionality. We present a general method of gauging dimensionality in high dimensional nonlinear systems subject to noise, which could be applied to many systems in the context of neuromorphic computing. Our work will inform future implementations of spatially multiplexed VCSEL PNNs.

	\end{abstract}
	
	\maketitle

\section{Introduction \label{sec:Intro}}

Artificial neural networks (ANNs) have become a ubiquitous computing technique. Indeed, due to their flexibility and high performance, they have revolutionized many fields ranging from natural language processing and object recognition \cite{goodfellow2016deep} to self driving vehicles\cite{bojarski2016end}. ANNs are fundamentally different from classical computers, in that they process information in a fully parallel manner. Thus, there has been growing interest in developing fully parallel hardware  to enable efficient implementations \cite{jouppi2018domain,dally2020domain}. Among these, photonics has been heralded as a promising platform in terms of scalability\cite{dinc2020optical,rafayelyan2020large,moughames2020three}, speed\cite{shen2017deep,brunner2013parallel}, energy efficiency \cite{miller2017attojoule} and parallel information processing\cite{brunner2013parallel}.\\

Reservoir computing (RC)\cite{jaeger2004harnessing,van2017advances} is a simple, efficient and yet high performance ANN concept where only the output weights are trained. Thus, RC can be implemented to leverage the computational power of many existing physical systems\cite{tanaka2019recent}, which makes it a relevant benchmark architecture that can be used to consistently gauge hardware performance. Ultimately, for  photonic neural networks (PNNs) to be truly competitive, they require high performance, efficiency, speed and scalability. Moreover, in situ training techniques should be implemented to avoid speed bottlenecks and to reduce the reliance on an auxiliary high performance computer \cite{brunner2021competitive}. Semiconductor lasers have emerged as major candidates to implement PNNs due to their ultrafast modulation rates and complex dynamics\cite{brunner2013parallel}. Among these, vertical-cavity surface-emitting laser (VCSELs) are of particular interest because of their efficiency, speed, intrinsic non linearity and the maturity of their CMOS-compatible fabrication process\cite{muller20111550,vatin2018enhanced}. In\cite{porte2021complete}, RC is implemented via the spatial multiplexing of modes on the surface of a large area VCSEL (LA-VCSEL). This new approach allows for a truly parallel and autonomous network where the role of the external computer is greatly minimized via the use of hardware learning rules. This implementation differs from the popular so-called time-multiplexed approach\cite{appeltant2011information,brunner2019photonic}, where information is still processed sequentially rather than in a parallel way and a heavy involvement of an external computer is the usual technique to interface and to create the reservoir state. \\

For the use of LA-VCSELs in PNNs, efforts must be expended in characterising these devices and their performance dependence on key physical parameters. Our investigation links these physical parameters to generic computational metrics, namely consistency and dimensionality. We find that injection locking conditions yield the best performance for our benchmark classification task (3-bit header recognition), reaching a $1.5 \%$ error rate. In addition, biasing the LA-VCSEL at higher currents improved performance due to a stronger non-linear response of our device. Consistency was measured to be above $99 \%$ highlighting the robustness of our system. Lastly, the dimensionality of our device was measured under several conditions and we find a correlation between higher dimensionality and better performance.

\section{Working Principle}\label{principle}

\begin{figure}[h]
	\begin{center}
		\includegraphics[width=1\linewidth]{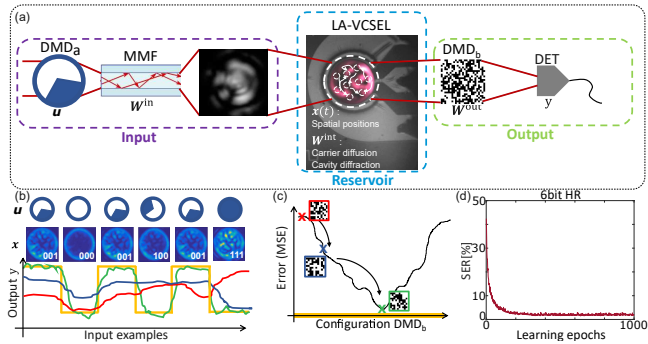}  
		\caption{(a) Working principle of the LA-VCSEL spatially multiplexed reservoir.
			(b) Input information $\mathbf{u}$ and the subsequent LA-VCSEL response for 3-bit binary headers.
			The graph shows the target output $y^{\text{target}}$ (yellow) for classifying header 001 and different reservoir outputs $y^{\text{out}}$ of decreasing mean square error (MSE) (red, blue and green).
			(c) Schematic illustration of the error landscape, showing the MSE as a function of the output weights configuration.
			The outlined (red, blue and green) Boolean matrices correspond to the output weights giving the output from (b).
			(d) Representative performance of the PNN on a 6-bit header recognition task.}
		\label{fig:SETUP}
	\end{center}
\end{figure}

The PNN implemented here is similar to the one presented in \cite{porte2021complete}. Like a conventional RC, it is divided into three sections: an input layer; a reservoir; and an output layer. The working principle of our system is shown in Fig.~\ref{fig:SETUP}(a).

The input layer is built using an continuously tunable external cavity laser (ECL, Toptica CTL 950), a digital micro-mirror device (DMD,Vialux XGA 0.7" V4100) and a multimode fibre (MMF, Thorlabs M42L01). The ECL beam is colimated and illuminates $\text{DMD}_\text{a}$. The mirrors on a DMD can flip between two positions, which allows us to display Boolean images that constitute our input information $\mathbf{u}$ (blue parts are on, white parts are off). Each image is displayed on $\text{DMD}_\text{a}$ for $200~ \mu \text{s}$ resulting in an injection frame rate of $5~\text{kHz}$, which is orders of magnitude slower than the intrinsic time scales of the VCSEL ($\sim \text{GHz}$), meaning that we use the device in its steady state. The spatially encoded input information is then sent through a MMF of $50~\mu \text{m}$ in diameter, which passively implements our input weights $\mathbf{W^{\text{in}}}$ via the MMF's complex transmission matrix. In practise, this is done by imaging the spatial pattern on $\text{DMD}_\text{a}$ through the MMF. The input information consists of sequences of binary pie-shaped headers that are split into several parts according to the number of bits of information that is displayed, Fig.~\ref{fig:SETUP}(a) shows a 3-bit header. Moreover, the input patterns exhibit an outer ring that is constantly in the "ON" state. The ring injects a DC-locking signal that is used to injection-lock the device to ensure its stability. The ring's thickness is the difference between the inner radius and the outer one, which is set by the size of the colimated injection beam on $\text{DMD}_\text{a}$. The thicker the ring, the more power is continuously injected to lock the LA-VCSEL, but the less power is used to encode the input information. The general injection locking aspect is explained in more detail in Section \ref{sec:bias}.\\

The nearfield output of the MMF, $\mathbf{W}^\text{in}\mathbf{u}$ is optically injected into the LA-VCSEL. This can be realized by imaging the MMF output facet's nearfield on the surface of the device, and here we use a LA-VCSEL with an aperture of $\sim 50~ \mu \text{m}$ and a threshold current $I_{\text{th}}=20 ~\text{mA}$. Our device was fabricated in a university grade clean-room following a process which has been optimized for high bandwidth and energy efficiency \cite{haghighi2019power,haghighi202040}. The VCSEL structure is the same as the one used in \cite{porte2021complete}. It is grown epitaxially and comprises a half-wavelength ($\lambda / 2$) cavity and two distributed Bragg reflectors (DBRs). The cavity hosts 5 InGaAs quantum wells, and top and bottom DBRs are respectively made of periodically alternating $Al_{x}Ga_{1-x}As$ layers. The top DBR is a 14.5 period p-doped structure with
$x=0.1$ and $x=0.92$, while the bottom DBR is an n-doped 37-periodic structure alternating between $x=0.05$ and $x=0.92$. Finally, the circular aperture of the VCSEL was defined through oxidization of a Al-rich central layer with an oxidization length of $\sim 9~\mu\text{m}$. In this setup, the components of the reservoir, that is to say non-linear nodes and the connections between them, are fully implemented by the physical properties and dynamics of the LA-VCSEL. Nodes are spatial positions on its surface, and the coupling between these neurons is taken care of via intrinsic physical processes. Namely, carrier diffusion in the semiconductor medium creates a Gaussian-like local coupling, while diffraction of the optical field inside the laser cavity creates a complex global coupling. The LA-VCSEL takes the input information and transforms it in a complex non-linear way according to the dynamics of optical injection. This process produces the reservoir state $\mathbf{x}$, shown in Fig.~\ref{fig:SETUP}(a) and Fig.~\ref{fig:SETUP}(b). Moreover, in Fig.~\ref{fig:SETUP}(b), we can see how the reservoir responds to different input information (different 3-bit headers). Each response is complex, non-linear, and encodes significantly modified responses for the different input information. These differences explain, in a simple way, why the system is able to learn, i.e. why it is possible to find a configuration of output weights that solve a certain task such as pattern classification.\\

The final component of our PNN is its programmable output layer in order to physically implement morphism and learning. The VCSEL's near field is imaged onto a second DMD ($\text{DMD}_\text{b}$), whose surface of this DMD is in turn imaged onto a large area photo-detector (DET, Thorlabs PM100A, S150C). We rely on $\text{DMD}_\text{b}$ to implement our output weights $\mathbf{W}^{\text{out}}$. The VCSEL itself is a spatially continuous medium, whereas the DMD is a discrete matrix of pixels. Therefore, by imaging onto $\text{DMD}_\text{b}$, we sample the VCSEL's surface with the pixels of the DMD, and setting the magnification allows us to tune the sampling rate (number of neurons). Here, we implement around $\text{n}=350$ fully parallel neurons. The mirrors on a DMD can flip between two position, one of which diverts the corresponding optical signal towards the DET, i.e.,  giving us Boolean readout weights. By choosing the right configuration of output mirrors, we can tune the spatial positions of the LA-VCSEL that contribute to the optical power detected at the DET and therefore train the output $\mathbf{y}$ of the reservoir.\\

When training the network, a random matrix of output weights is loaded onto $\text{DMD}_\text{b}$. The output $\mathbf{y}$ is recorded for a sequence of $T$ images (training sequence) as shown in Fig.~\ref{fig:SETUP}(b) and Fig.~\ref{fig:SETUP}(c)-~$T$ is therefore the batch size. The target output $\mathbf{y^{\text{target}}}$ is known for every input pattern belonging to the training set. After each learning epoch $k$, (a run through all the input images), $\mathbf{y}$ is recorded and a normalized mean square error (NMSE) is calculated: 

\begin{equation}
	\epsilon_{k} =\frac{1}{T}\sum_{t=1}^{T}(\mathbf{y}_{k}(t) - \mathbf{y}^{\text{target}}(t))^2.
\end{equation}

\noindent Training is realised via a simple, yet effective evolutionary algorithm presented in \cite{bueno2018reinforcement,andreoli2020boolean}.
A single or numerous mirrors located at randomly chosen positions are flipped at the transition between epochs $k$ and $k+1$.
If the change results in a lower error, i.e. $\epsilon_{k+1} < \epsilon_{k}$, it is kept, otherwise the output weights are reset to the configuration at epoch $k$ as shown in Fig.~\ref{fig:SETUP}(b) and Fig.~\ref{fig:SETUP}(c). Finally, Fig.~\ref{fig:SETUP}(d) shows a representative learning curve for a 6-bit header recognition task, demonstrating that our PNN can also be applied to significantly more challenging tasks. In this task the VCSEL has to differentiate between $2^6=64$ classes. After learning, which takes around a minute for each class, the system reaches a symbol error rate (SER) of around $1.5 \%$ averaged accross all 64 classes.

\section{Computing Performance Analysis}

In order to evaluate the impact of several parameters on learning we use a 3-bit header recognition task (illustrated in Fig.~\ref{fig:SETUP}(b)) as our benchmark for convenience and speed. The goal in this task is to recognize every input pattern individually among the $2^3=8$ images. Our computational performance metric will be the NMSE error and the SER. The relevant parameters studied in this work are the injection wavelength $\lambda_{\text{inj}}$, the injection power ratio PR, the LA-VCSEL's bias current $I_{\text{bias}}$ and the fraction of the injection power assigned to the DC locking ring.

\subsection{Physical parameters}
\subsubsection{Injection wavelength and power}
\begin{figure}[h!]
	\begin{center}
		\includegraphics[width=1\linewidth]{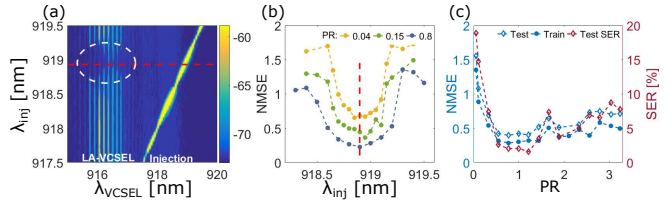}
		\caption{(a) Injection locking of the VCSEL by an external drive laser (power ratio, $\text{PR} =0.8$), the red dotted line shows the resonance at $\lambda_{\text{inj}} = 918.9~ \text{nm}$ and the white dotted circle shows the locking range. (b) Performance (NMSE) vs injection wavelength for different injection power ratios (PR), highlighting that the best performance is reached for the injection locking conditions. (c) NMSE (left axis) and SER (right axis) as a function of PR (taken at the optimal wavelength detuning condition, $\lambda = 918.9 ~\text{nm}$). For all measurements, the bias current was $50\%$ above threshold, at this current the VCSEL emits $\approx 3.6~\text{mW}$ of optical power.}
		\label{fig:lambdascan}
	\end{center}
\end{figure}
First we show how the injection locking conditions impact our system. For reference, the VCSEL's output power was $3.6 ~ \text{mW}$ when biased $50$\% above threshold, $I_{\text{bias}}=1.5I_{\text{th}}$, with $I_{\text{th}}=20~\text{mA}$. Fig.~\ref{fig:lambdascan}(a) shows how the LA-VCSEL's free running modes react to an external drive laser. We scan the injection wavelength continuously until a resonance condition is met at $\lambda_{\text{inj}} = 918.9~ \text{nm}$. At this wavelength, the VCSEL's free running modes are suppressed by roughly $10~\text{dB}$ and its emission wavelength is shifted to that of the injection laser. This phenomenon is called injection locking. The mechanisms of optical injection for multimode LA-VCSELs have been extensively studied in \cite{ackemann1999spatial,ackemann2000patterns}. Here, we study the impact of injection locking on computational performance.\\

Fig.~\ref{fig:lambdascan}(b) highlights the importance of the injection wavelength. We see a clear and consistent best performance basin around the resonance wavelength ($\text{NMSE} =0.2$). Besides the spectral resonance, we also find that the performance degrades with a lower injection power ratio ($\text{PR}=P_{\text{inj}}/P_{\text{VCSEL}}$), a trend which is confirmed by Fig.~\ref{fig:lambdascan}(c). There, we see that the performance increases until $\text{PR}\sim1$ where it saturates and then starts to degrade again for higher PRs. The best performance is reached when the device is fully injection locked, yet apparently before overly intense input has quenched its nonlinearity. In an intuitive sense, for classification tasks, we want the LA-VCSEL's response to be sensitive to changes in the input information yet still produce reliable results, and these conditions are met under injection locking. Our results are in line with the ones presented in\cite{bueno2017conditions,bueno2021comprehensive} for an edge-emitting laser and a VCSEL-based time delay reservoir respectively.

\subsubsection{Bias current and DC-locking signal strength \label{sec:bias}}

The next physical parameter that is relevant for our system is the bias current $I_{\text{bias}}$. It controls the response of the device by acting on the carrier-concentration distribution in the semiconductor medium, which can impact the non-linearity of the device and the modal configuration of its free-running emission. Thus, we measured the performance of our PNN for different bias currents above threshold, with $I_{\text{th}}=20 ~\text{mA}$. In addition, for each bias current, we performed a scan of the outer ring thickness, and hence for different ratios between the DC injection power and the optical power used to encode the input information.\\

\begin{figure}[h!]
	\begin{center}
		\includegraphics[width=1\linewidth]{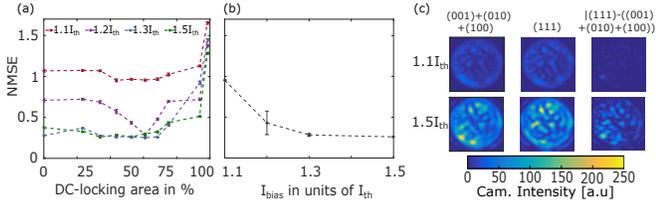}
		\caption{(a) NMSE as a function of the DC-locking area for different bias currents, $I_{\text{th}} = 20~m\text{A}$. At $0$\%, there is no DC-injection locking signal, while at close to $100$\% no information is injected into the system. (b) NMSE as a function of the bias current averaged for DC-locking areas between $30\%$ and $70\%$. (c) Left: summed images of the VCSEL's responses to the input headers (001), (010) and (100). Middle: image of the VCSEL's response to input header (111). Right: absolute value of the difference between the left and middle images, highlighting that the overall system's response is more nonlinear at a higher bias current.}
		\label{fig:Ibiasscan}
	\end{center}
\end{figure}

Fig.~\ref{fig:Ibiasscan}(a) shows how the performance varies as a function of the DC-locking area for different bias currents. Generally, the best performance is achieved when the DC-locking area is between $30\%$ and $70 \%$ of the total input area. For the highest bias current, $I_{\text{bias}}=1.5I_{\text{th}}$ we see that this outer ring may not be needed, at least when we operate the LA-VCSEL PNN in its steady state as done here.\\

Fig.~\ref{fig:Ibiasscan}(b) shows that the best performance is reached for $I_{\text{bias}}=1.5I_{\text{th}}$, which is substantiated by Fig.~\ref{fig:Ibiasscan}(c), where we recorded the response of the VCSEL to headers $001$, $010$ and $100$ with a camera. We then summed these responses and compared them to the real response of the VCSEL to header $111$. The difference shown in the third column cannot be solely attributed to a stronger VCSEL nonlinear response. Indeed, optical detection, in the way we do it here, is itself nonlinear because we measure optical intensities with a camera. Yet, this comparison provides a sense of the total nonlinear spatial response of the system comprising the VCSEL and optical detection. It shows a significantly greater difference for the two subtracted images at $I_{\text{bias}}=1.5I_{\text{th}}$ than at $I_{\text{bias}}=1.1I_{\text{th}}$, suggesting that the system's response becomes more nonlinear and providing a potential explanation for the gain in performance, going from an NMSE of $1$ at $1.1I_{\text{th}}$  to $0.2$ at $1.5I_{\text{th}}$.

\subsection{Computational metrics}

In the previous section, we studied the impact of injection wavelength, power and bias current on the computational performance of our PNN, using digital header recognition as a benchmark task. Here, we will focus on computational metrics, namely dimensionality and consistency, and how these vary with the physical parameters previously studied. These metrics are generic in nature, and as such, can be measured for different hardware platforms, which consequently allows for an essential hardware-agnostic comparison. In the case of dimensionality, we attempt to establish a general way to gauge an analog hardware ANN's dimensionality using principal component analysis, which is non-trivial due to the presence of noise in such systems.

\subsubsection{Consistency Analysis}

Consistency is defined as the ability of a system to respond in the same way when subjected to the same input information\cite{kanno2012consistency,nakayama2016laser}. Consistency is a fundamental property of dynamical systems and is of high relevance when considering new devices or platforms for neuromorphic hardware\cite{bueno2017conditions}. Indeed, a system that is not consistent would not be able to learn since its responses would not be reproducible. In practice, the analysis is done by injecting the system with the same input information several times, recording its response, and then computing the cross-correlation matrix of these responses. Due to the symmetry of correlation matrices, the consistency is the mean of the upper diagonal part of said matrix. As an input, we chose a random sequence of binary headers comprising 1000 images.\\

\begin{figure}[h!]
	\begin{center}
		\includegraphics[width=1\linewidth]{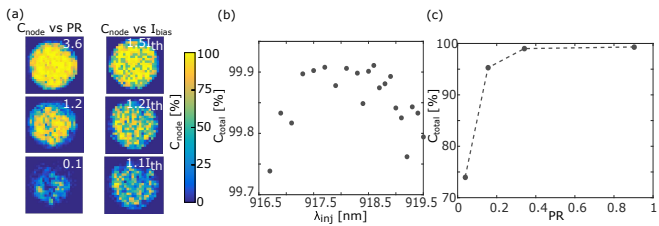}
		\caption{(a) 2D distribution of the per-node consistency $\text{C}_\text{node}$ for different injection power ratios (left) and bias currents (right). For the left column $I_{\text{bias}}=1.5I_{\text{th}}$ was constant, for the left column $\text{PR} \sim 1$ was constant. (b) Total system consistency  $\text{C}_\text{total}$ as a function of the injection wavelength. (c) Total system consistency  $\text{C}_\text{total}$ as a function of the injection power ratio.}
		\label{fig:Consistency}
	\end{center}
\end{figure}

In Fig.~\ref{fig:Consistency}, we show how the total consistency $\text{C}_\text{total}$ (all mirrors on $\text{DMD}_{\text{b}}$ are switched on simultaneously) as well as the individual node-resolved consistency $\text{C}_\text{node}$ (every mirror is switched on individually) vary as a function of the parameters studied in previous sections. Figure~\ref{fig:Consistency}(a) shows how the node-resolved consistency varies as a function of PR (left side) and $I_{\text{bias}}$ (right side). First, the bias current is fixed at its optimal value obtained in  Fig.~\ref{fig:Ibiasscan}(b), i.e $I_{\text{bias}}=1.5I_{\text{th}}$, and the injection power is swept. Then, the injection power is fixed so as to have the optimal power ratio obtained from Fig.~\ref{fig:lambdascan}(c), i.e $\text{PR} \sim 1$.  The overall trend is that the consistency of individual nodes increases when $P_{\text{inj}}$ and $I_{\text{th}}$ increase, going from a mean value of $16 \%$ to $60 \%$ then $88 \%$ for $\text{PR}=0.1,1.2,~\text{and}~3.6$ and from $30\%$ to $50\%$ and $70 \%$ for $I_{\text{bias}}=1.1,1.2,~\text{and}~1.3I_{\text{th}}$.
 \\  

Fig.~\ref{fig:Consistency}(b) shows the overall system consistency and its dependence on the injection wavelength $\lambda_{\text{inj}}$.  Overall, the total consistency $\text{C}_\text{total}$ remains above $99 \%$. As a consequence, we cannot solely attribute the drop in performance as a function of $\lambda_{\text{inj}}$ seen in Fig.~\ref{fig:lambdascan}(b) to a consistency issue. Finally, Fig.~\ref{fig:Consistency}(c) shows how the consistency varies as a function of the injection power ratio. We see that it saturates at a ratio $\text{PR}\sim 0.35$, at this power level, the NMSE drops to $\sim 0.5$ from $\sim 0.9$ at $\text{PR}\sim0.1$, as shown in Fig.~\ref{fig:lambdascan}(c).\\

The main detrimental factor decreasing consistency in our system, and creating a substantial difference between the node-resolved and global consistencies, is noise. Neglecting the noise of the very stable injection laser, we are left with two noise sources. First, spontaneous emission from the VCSEL acts as an uncorrelated noise source that is added to the response of individual nodes without correlation. Second, the photodiode we use for detection presents different types of noise, namely, thermal noise, shot noise, and dark current noise \cite{hui2019introduction}. Yet, the photodiode was rated to measure signals whose noise level is significantly smaller than $0.1~\text{nW}$. The lowest signal we send (dimmest individual node) is around $0.1~\mu\text{W}$, and this difference of at least three order of magnitude between detection noise level and signal means that the setup is not detection noise limited. Regarding spontaneous emission noise, it is more prevalent when close to the LA-VCSEL's threshold, which explains the increase in node-resolved consistency seen in Fig.~\ref{fig:Consistency}(a) when increasing the bias current. Furthermore, increasing the injection power ratio PR yields better locking conditions which stabilize the device and yield more reproducible dynamical responses to input information, reducing the impact of spontaneous emission noise and increasing the consistency as shown also in Fig.~\ref{fig:Consistency}(a). Lastly, because spontaneous emission is uncorrelated, it is efficiently averaged out when  measuring the total consistency \cite{semenova2019fundamental}, which explains the significant drop in relative noise amplitude for node-resolved and total consistency measurements seen in Fig.~\ref{fig:Consistency}.

\subsubsection{Dimensionality Analysis}

The last parameter we study is the dimensionality of our PNN. When selecting a potential hardware candidate for implementing a PNN or an analog NN in general, a method to systematically characterize its dimensionality is essential to provide system-independent comparability. Here, we present a method to achieve such a goal. In this direction, previous work gauged the dimensionality of dynamical systems in terms of computation based on an expansion of the system's responses in a space of orthogonal functions \cite{dambre2012information}. However, such an expansion only provides the correct result in the absence of noise. Here, we considerably expand the validity of dimensionality estimation using a method that is in no way limited to our system and solely relies on injecting input information and recording responses of individual nodes separately.\\

First, a random sequence of input binary headers of $N_{\text{bits}}$ is generated. This sequence is $\text{T}=1000$ images long, each image representing a single time step. Each mirror (weight) on $\text{DMD}_\text{b}$ is set to its "ON" state individually, and the corresponding node's response to the input information is recorded. We then define the state-collect matrix, $\mathbf{M}$ as a matrix whose columns are the individual node responses $\text{N(t)}$. Considering the $\text{n}=350$ we defined for our PNN, the matrix $\mathbf{M}$ will therefore be $\text{T} \times \text{n} $ in size:

\begin{equation}\mathbf{M}=\left[\begin{array}{ccc}
		{N_{1}({1})} & \cdots & {N_{n=350}({1})} \\
		\vdots & \ddots & \vdots \\
		{N_{1}(T={1000})} & \cdots & {N_{n=350}(T={1000})}
	\end{array}\right].\end{equation}

Computing the dimensionality of $\mathbf{M}$ would allow us to gauge the dimensionality expansion of the input data performed by our LA-VCSEL. This expansion, which results in the mapping of input data to higher dimension space forms the basis of ANNs, and explains why they can produce non-trivial decision boundaries which solve complex computational tasks. Nonlinearity is needed for such dimensionality expansions, which explains why neurons present nonlinear activation functions in ANNs. Indeed, tasks which are not solvable in the low dimensional input space can benefit from a nonlinear mapping onto a higher dimensional space \cite{cover1965geometrical,appeltant2011information}.
Therefore, measuring the dimensionality of the LA-VCSEL's nonlinear transformation is crucial to understanding its computational power. In a noiseless scenario, such a task would be simple and indeed, one would just need to calculate the rank of $\mathbf{M}$, which would give us the number of linearly independent node responses, and therefore, the dimensionality of the state space.
Yet, noise will add a random modulation on top of each individual node response, such that even two perfectly linearly dependant nodes will become partially linearly independent after noise is added. Sources of noise in our system include the injection laser, the LA-VCSEL and the photodetector, and it is a general feature of analog neural networks \cite{semenova2019fundamental,andreoli2020boolean}. Although in this case, noise increased the dimensionality, it cannot be leveraged for computation because its contributions are random and not consistent. As a consequence, to carry out a dimensionality analysis for noisy data one has to rely on methods which deal with correlations rather than pure linear dependencies. Principal component analysis (PCA)\cite{wold1987principal,hotelling1933analysis,pearson1901liii} is such a method.\\

PCA applies a dimensionality analysis by giving orthogonal principal components along which the variance in the data is distributed and assigns a weight to these principal components. If columns of $\mathbf{M}$ are predominantly linearly dependant, they will get mapped onto a small number of principal components, which in turn will account for nearly all of the variance in the data. However, because noise adds variance in the data, some principal components will be noise-dominated therefore artificially adding dimensionality to our dataset. The crucial challenge facing us here, is to identify the right number of principal components that explain real variance in the data, while excluding those that  predominantly explain noise. We would thus like a systematic way of excluding the principal components that are noise dominated, and carrying out this analysis will give us an estimate for the rank of $\mathbf{M}$.\\

In order to do so, we first compute $\mathbf{\Sigma}$, the covariance matrix of $\mathbf{M}$:

\begin{equation}\mathbf{\Sigma}=\operatorname{cov}(\mathbf{M})=\left[\begin{array}{ccc}
		\operatorname{cov}\left(N_{1}(t), N_{1}(t)\right) & \cdots & \operatorname{cov}\left(N_{1}(t), N_{n}(t)\right) \\
		\vdots & \ddots & \vdots \\
		\operatorname{cov}\left(N_{n}(t), N_{1}(t)\right) & \cdots & \operatorname{cov}\left(N_{n}(t), N_{n}(t)\right)
	\end{array}\right].\end{equation}

\noindent We then carry out a singular value decomposition (SVD) of $\mathbf{\Sigma}$. The SVD algorithm finds the  matrices $\mathbf{U, S, V}$  that satisfy:  

\begin{equation}\mathbf{\Sigma}=\mathbf{U S} \mathbf{V}^{T},\end{equation}

\noindent where $\mathbf{S}$ is a diagonal matrix whose entries $\Lambda_{\text{k}} , \text{k} \in\{1, \dots ,\text{n}=350\}$, are the eigenvalues of $\mathbf{\Sigma}$ and its eigenvectors (principal components) are the columns of $\mathbf{U}$. The eigenvalues represent the amount of variance explained by each principal component. Since $\mathbf{\Sigma}$ is symmetric and real valued, $\mathbf{U}=\mathbf{V}$, and $\mathbf{U}$ is real valued. $\mathbf{U, S, V}$ are all square matrices of size n$\times$n because $\mathbf{\Sigma}$ is square. As stated previously, we would like a systematic way of identifying noise dominated principal components. In \cite{malinowski1977theory}, such a criterion is presented. This is possible under the assumption that said noise is Gaussian in origin and distributed with a constant standard deviation for all elements (neurons) and at all times. The author gives a statistical indicator $I$, named the factor indicator function, which is a function of the eigenvalues $\Lambda_{\text{k}}$:

\begin{equation} \operatorname{I}(k)=\frac{\left(\sum_{i=k+1}^{n} \Lambda_{i}^{}/{T(n-k)}\right)^{1 / 2}}{(n-k)^{2}}.\end{equation}

\noindent$I(k)$ is related to the difference between the noisy experimental data and the noiseless data. A more in depth explanation is given in \cite{malinowski1977theory} and in \cite{turner2006noise}, where this method was applied to remove noise and reduce the dimensionality of an atmospheric emitted radiance dataset.  Locating the minimum of $I(k)$ at $k_{\text{min}}$ yields  then the number of principal components representing \emph{meaningful} variance in the data, while minimizing the influence of noise, giving us therefore an estimate of $\mathbf{M}$'s rank.\\

\begin{figure}[h]
	\begin{center}
		\includegraphics[scale=0.8]{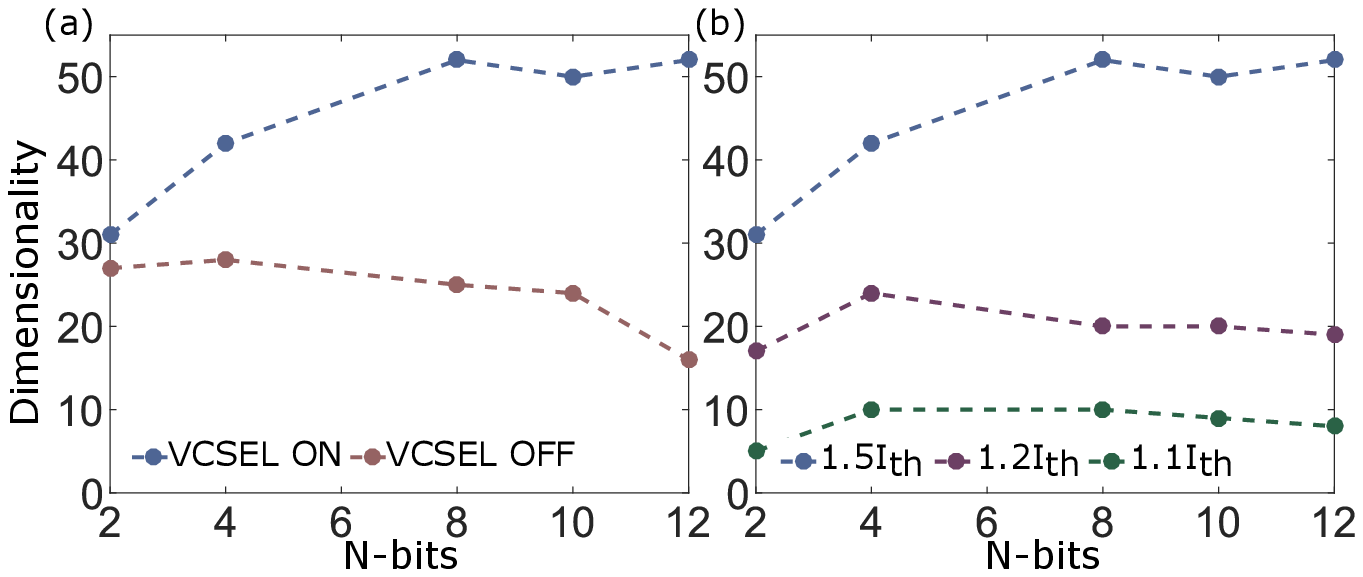}
		\caption{Dimensionality measurements for different input bit-numbers and PR$\sim 0.8$. (a) Dimensionality of the system with the VCSEL ON and OFF. The VCSEL expands the dimensionality of the input highlighting the non-linearity of the device.  (b) Dimensionality for different bias currents, the dimensionality increases with $I_{\text{bias}}$ highlighting the stronger non linear nature of the system at higher bias currents.}
		\label{fig:Dimesionality}
	\end{center}
\end{figure}

Fig.~\ref{fig:Dimesionality} shows the results of this PCA method on our dataset containing the node responses for a different number of input dimension N-bits. In Fig.~\ref{fig:Dimesionality}(a), the dimensionality is measured and compared with the VCSEL switched ON and OFF. The notable difference between the two configurations clearly shows that our device significantly increases the dimensionality of the input data.  This increase in dimensionality is in general, but not always, helpful for computation. Indeed, when switching the VCSEL off, we found that our system was not able to learn any task. Figure~\ref{fig:Dimesionality}(b) shows how the bias current impacts the dimensionality of our system at a constant injection power ratio $\text{PR} = 0.8$, and results are generally in-line with  Fig.~\ref{fig:Ibiasscan}. There is a clear correlation between the increase in dimensionality and better performance for higher bias currents. Moreover, Fig.~\ref{fig:Ibiasscan}(c) shows that the response of the system is overall more nonlinear. This stronger non linearity is consistent with an increased dimensionality at higher bias currents.

Finally, one should keep in mind that several assumptions and approximations were made to obtain these dimensionality numbers. As such, they should not be interpreted as absolute values, but rather in comparison with each other. Common to all data in Fig.~\ref{fig:Dimesionality} is that the dimensionality saturates and in some cases even drops when the number of input dimensions exceeds 10 bits. We associate this trend to the reduced optical intensity that is injected into the device per bit, and the resulting smaller modifications created by different bit configurations because increasingly smaller regions of the input area on $\text{DMD}_\text{a}$ are used to encode each bit.\\

We see that under the best conditions, $I_\text{bias}=1.5I_\text{th}$ we reach a dimensionality of $50$, whereas we implement $350$ weights. We can compare this number to the number of modes supported by the VCSEL, which can be estimated by taking the ratio between the areas of a typical central speckle and the area of the VCSEL. We find that the VCSEL supports at least $80$ modes. We can then compare the size of these speckles $\sim 5.6~\mu \text{m}$ to the size of one mirror on $\text{DMD}_\text{b}$, and taking into account the magnification we get sampling factor of $\sim1.6$, every speckle is therefore imaged onto $1.6$ mirrors. We then find that $50 = 80/1.6$ making our estimation of the dimensionality via PCA roughly inline with the physics of the device.\\
Lastly, our measurement is blind to phase and polarization dynamics. Indeed, we do not use phase or polarization to encode information, nor do we consider it when training or measuring dimensionality hence, we systematically underestimate the dimensionality of our device. Capturing its full complexity would require more elaborate encoding and detection schemes.

\subsection{Conclusion}

We studied the impact of certain physical parameters, namely, injection wavelength, injection power, and bias current on performance in a spatially multiplexed LA-VCSEL-based photonic neural network for the first time. We showed that for our classification task, the best performance is achieved when injection locking conditions are met in terms of injection wavelength and power. Moreover, we investigated the impact of the bias current on performance, and showed that a higher bias current resulted in better performance as well as a stronger system nonlinearity. Although our study was limited to a single benchmark task, i.e. 3-bit header recognition, the general conclusions drawn in this work should still be applicable to other classification tasks, especially when considering the impact of injection wavelength and power. Indeed, our work is inline with previous studies conducted on VCSEL and edge-emitting laser based PNNs. Moreover, in the case of hardware NNs fine-tuning of physical parameters is needed and this work provides a useful analysis and phenomenological explanations for the adequate range of physical parameters that was measured. We then studied the impact of these physical parameters on useful computational metrics for neuromorphic hardware, namely: consistency and dimensionality. We measured a high total system consistency (above $99 \%$), and studied how this consistency changed with different physical parameters. Lastly, we presented a general method for gauging the dimensionality of a hardware system based on principal component analysis under the influence of noise. We were able to establish some correlations between higher dimensionality and better performance under certain conditions. This confirms the consistency of our dimensionality estimation with previous measurements. Our work is of high relevance as it prepares using LA-VCSELs as building blocks of future, more complex PNN structures such as VCSEL arrays \cite{heuser2019development,heuser2020developing}. Finally, we presented a simple, efficient and general method of estimating dimensionality that could be applied to many systems.

\section*{Funding}
The authors acknowledge the support of the Region Bourgogne Franche-Comt\'{e}. 
This work was supported by the EUR EIPHI program (Contract No. ANR-17-EURE- 0002), by the Volkswagen Foundation (NeuroQNet I\&II), by the French Investissements d’Avenir program, project ISITE-BFC (contract ANR-15-IDEX-03), partly by the french RENATECH network and its FEMTO-ST technological facility, and by the German Research Foundation (via SFB 787),and by the European Union’s Horizon 2020 research and innovation program under the Marie Skłodowska-Curie grant agreement No 713694 (MULTIPLY) and 860830 (POST DIGITAL).

\section*{Disclosures}
The authors declare no conflicts of interest.

\bibliography{biblio}

\end{document}